\documentclass[preprint,times,twocolumn]{aastex6}
\usepackage[fleqn]{amsmath}
\usepackage{natbib}
\usepackage{enumitem}
\usepackage{booktabs}
\usepackage{graphicx}
\usepackage{hhline}

\begin{document}

\title{The \textbf{BRAVE} Program - I: Improved Bulge Stellar Velocity Dispersion Estimates for a Sample of Active Galaxies}

\author{Merida Batiste\altaffilmark{1}, Misty C. Bentz\altaffilmark{1}, Emily R. Manne-Nicholas\altaffilmark{1}, Christopher A. Onken\altaffilmark{2}, and Matthew A. Bershady\altaffilmark{3}}
\begin{center}
\altaffiltext{1}{Department of Physics \& Astronomy, Georgia State University, 25 Park Place, Atlanta, GA 30303, USA; batiste@astro.gsu.edu}
\altaffiltext{2}{Research School of Astronomy \& Astrophysics, The Australian National University, Canberra, ACT 2611, Australia}
\altaffiltext{3}{Department of Astronomy, University of Wisconsin, 475 N. Charter St., Madison, WI 53706, USA}
\end{center}

\begin{abstract}
We present new bulge stellar velocity dispersion measurements for 10 active galaxies with secure $M_{BH}$ determinations from reverberation-mapping. These new velocity dispersion measurements are based on spatially resolved kinematics from integral-field (IFU) spectroscopy. In all but one case, the field of view of the IFU extends beyond the effective radius of the galaxy, and in the case of Mrk 79 the field of view extends to almost one half the effective radius. This combination of spatial resolution and field of view allows for secure determinations of stellar velocity dispersion within the effective radius for all 10 target galaxies. Spatially resolved maps of the first (V) and second ($\sigma_{\star}$) moments of the line-of-sight velocity distribution (LOSVD) indicate the presence of kinematic substructure in most cases. In future projects we plan to explore methods of correcting for the effects of kinematic substructure in the derived bulge stellar velocity dispersion measurements.

\end{abstract}

\section{Introduction} \label{sec:intro}
Over the last several decades observational studies have revealed a fundamental connection between the formation and evolution of galaxies and that of their central super-massive black holes (BHs) (see \citealt{Kormendy13} for a review). This connection is exemplified by the existence of several scaling relations between the mass of the central BH, $M_{BH}$, and properties of the host galaxy, including bulge stellar velocity dispersion, $\sigma_{\star}$ \citep{Ferrarese00,Gebhardt00,Gultekin09}. 

These scaling relations are determined based on a set of well constrained $M_{BH}$ measurements, which have usually been made by modeling the spatially resolved stellar \citep{vandermarel98,Valluri04} or gas kinematics \citep{Macchetto97,Denbrok15} within the BH sphere of influence. This method requires high spatial resolution which limits its applicability to very nearby, and almost exclusively, quiescent galaxies. In active galactic nuclei (AGN) stellar dynamical modeling is impossible in all but a few cases \citep{Davies06,Onken07,Onken14}, due to the typically prohibitive distances, so reverberation mapping is used to determine $M_{BH}$ instead. Reverberation mapping exploits the variability of the AGN continuum; the response time of gas in the broad line region (BLR) to flux variations in the accretion disk is used to measure the size of the BLR ($R_{BLR}$). $M_{BH}$ can then be determined via the virial theorem, $M_{BH}=f\:\Delta\:V^{2}R_{BLR}/G$, where G is the gravitational constant, $\Delta\:V$ is measured from the width of the broad line, and $f$ is a scale factor that depends on the geometry of the BLR. Reverberation mapping is time resolution limited and is thus observationally intensive, but can in principle be applied to AGN at any distance. Consequently, AGN are the only tracers of $M_{BH}$ that can be used for studies over cosmological distances.

Reverberation mapping allows for accurate determination of the so-called virial product (VP), given by $M_{BH}/f$. The accuracy of $M_{BH}$ measurements are therefore dependent on the determination of $f$. Direct modeling of very high fidelity reverberation datasets (e.g., \citealt{Pancoast12}) can place constraints on the value of $f$ for individual objects, but is limited by our currently poor knowledge of the BLR and the simplicity of the assumptions inherent in the modeling. Furthermore, high fidelity datasets are only available for a handful of reverberation targets, so direct determination of $f$ for each object is not usually possible. Instead, $f$ is estimated by assuming that AGN and quiescent galaxies follow the same $M_{BH}-\sigma_{\star}$ relation \citep{Onken04}, and so can be found by determining the average multiplicative offset for $M_{BH}$ that is needed to bring the AGN relation into agreement with that of the quiescent galaxies. Ensuring that AGN black hole masses are accurately calibrated requires the assumption that active and quiescent galaxies do indeed follow the same relation (e.g., \citealt{Woo13}), and also depends on secure measurements of VP and $\sigma_{\star}$. 

Accurate estimation of $\sigma_{\star}$ for the bulge of a galaxy is complicated by a variety of considerations: There is no definitive evidence indicating the optimum aperture within which $\sigma_{\star}$ should be measured, in order to be most physically meaningful or to provide the tightest fit to the $M_{BH}-\sigma_{\star}$ relation. Values of $r_{e}/8$ (e.g. \citealt{Ferrarese00}) and $r_{e}$ (e.g. \citealt{Gebhardt00}) are frequently used, but $\sigma_{\star}$ estimates are not consistently determined within a specific radius and often not homogenized when taken from the literature. This problem is complicated by the obvious need for reliable determinations of $r_{e}$, which are often difficult to obtain. 

Similar to the issue of what fraction of $r_{e}$ to use is the unresolved question of how $\sigma_{\star}$ should be defined, and how to identify and remove contributions from contaminating substructure (and indeed whether or not it is necessary and appropriate to do so). It has been shown that the presence in disk galaxies of substructure such as bars and pseudo-bulges (characterized by a ``disky" bulge) may increase the measured $\sigma_{\star}$ (e.g. \citealt{Hu08,Graham11}), and the presence of a typical strong bar has been shown to increase scatter in $\sigma_{\star}$ by $\sim10\%$, which increase is strongly correlated with inclination \citep{Hartmann14}. Inclined disk galaxies present the added complication of disk light contaminating the bulge, which simulations indicate can increase $\sigma_{\star}$ by $10-25\%$, irrespective of aperture size \citep{Debattista13,Hartmann14}. This is a particular problem for maser galaxies, for which extremely accurate $M_{BH}$ determinations can be made, but which tend to be near edge-on (e.g. \citealt{Herrnstein05,Greene16} and references therein). Since the majority of $\sigma_{\star}$ determinations available in the literature come from long-slit spectroscopy or from single large fibers (as with the SDSS), the extent to which the contamination by kinematically distinct substructure can be accounted for is severely limited, and thus the actual uncertainties in reported values of $\sigma_{\star}$ are difficult to quantify.   

The \textbf{B}ig \textbf{R}everberation-mapped \textbf{A}GN \textbf{V}elocity dispersion \textbf{E}xamination (\textbf{BRAVE}) Program is a long term program intended to investigate and address these issues specifically for active galaxies. The first step in this process is to obtain spatially resolved kinematics within the bulges of active galaxies with secure $M_{BH}$ determinations. This can be done with integral-field spectroscopy, which allows individual spectra to be taken at points across a two dimensional field of view, from which spatially resolved kinematics maps can be generated. In this paper we present the first results of the \textbf{BRAVE} Program; spatially resolved kinematics for 10 reverberation-mapped active galaxies. For the initial findings presented here we report $\sigma_{\star}$ within $r_{e}$, determined as a weighted average of the values within that radius, without any corrections for the presence of substructure. It will be the goal of future work to determine the impact of kinematic substructure on these results, and any corrections that should be applied. Throughout this work we adopt a $\Lambda$CDM cosmology with $\Omega_{m}=0.3$, $\Omega_{\Lambda}=0.7$, and $H_{0}=70\:\mathrm{km\:s^{-1}\:Mpc^{-1}}$.

\section{Observations and Analysis} \label{sec:obs}
\subsection{HexPak} \label{subsec:hexpak}
\subsubsection{Observations} \label{subsub:hexpakobs}
Observations of eight AGN host galaxies were made using the HexPak integral-field unit \citep{hexpak} on the WIYN\footnote{The WIYN Observatory is a joint facility of the University of Wisconsin-Madison, Indiana University, the National Optical Astronomy Observatory and the University of Missouri.} 3.5m telescope at Kitt Peak National Observatory for four nights in April 2015, under NOAO program 2015A-0199. HexPak, which feeds into the Bench Spectrograph, is a PI instrument that was installed in late 2013 and has been available in shared use mode through NOAO since semester 2014B. It has a hexagonal field of view $40\farcs9 \times 35\farcs8$, consisting of 102 fibers of two different sizes. There is a central bundle of eighteen $0\farcs94$ fibers which subtends $6\arcsec$ on the sky, surrounded by a hexagonal array of 84 $2\farcs9$ fibers. In addition there are nine sky fibers arranged in an L-shape approximately $43\arcsec$ from the outer edge of the hexagon. This instrument is ideal for near face-on galaxies that are fairly large on the sky. The central bundle of fibers provides higher spatial resolution of the bright central region where S/N is not a problem, and the larger fibers provide higher S/N at the expense of spatial resolution in the lower surface-brightness outer regions of the galaxy.

We observed eight low inclination, late-type active galaxies, chosen from the database of galaxies with RM-based black hole mass estimates \citep{Bentz15}. Our sample was selected to consist of galaxies that are large enough on the sky to allow the bulge of the galaxy to be spatially resolved, while also having part of the disk included in the field of view in order to provide a large scale view of the dynamics. Targets were selected to provide a variety of substructure in the overall sample, since a long term goal of this project is to investigate the effects of substructure on $\sigma_{\star}$ measurements. Details of the target galaxies are shown in Table \ref{tab:gal_specs}, including redshifts as reported by NED\footnote{The NASA/IPAC Extragalactic Database (NED) is operated by the Jet Propulsion Laboratory, California Institute of Technology, under contract with the National Aeronautics and Space Administration.}. 

The 860 lines $\mathrm{mm}^{-1}$ grating blazed at $30.9^{\circ}$ was used in second order, targeting the wavelength range $4600-5600\mathrm{\AA}$ which contains the Mg\textit{b} absorption lines ($\lambda5167, 5173, 5184$\AA), with a spectral resolution of $2.02\mathrm{\AA}$. For each galaxy the array was centered on the AGN and observations were taken in 30 minute exposures. Total integration times were estimated based on the results of \citealt{Cortes06} who used the DensePak IFU to perform similar types of observations of galaxies in the Virgo Cluster. DensePak was the predecessor to HexPak, and fibers from that instrument were used in the construction of HexPak. The estimates of total integration time accounted for recent upgrades that have been made to the Bench Spectrograph, and were intended to obtain a $\mathrm{S/N}\sim30$ across the field of view. 

Observations were made during grey time and conditions during the course of observing were generally clear, though fog and precipitation prevented observing during the second half of the final night. However, moderate winds were present throughout. 
In order to minimize the effects of windshake on the data, significant efforts were made throughout the run to point the telescope in directions where it would be stable, and the pointing was checked and adjusted as necessary so that the AGN was centered on a single fiber before each new exposure. As a result of the wind we were unable to observe one of our primary targets, NGC 4748, and instead observed the secondary target Mrk 279. 

Total integration times for each galaxy are listed in Table \ref{tab:gal_specs}. Targets were observed at typical airmass of 1.3, and mean seeing was $\sim1\farcs5$. Spectrophotometric standards were observed throughout the night to facilitate telluric correction and flux calibration, and arc lamp spectra were taken at each pointing to facilitate accurate wavelength calibration. 

\begin{table*}
\begin{minipage}{\textwidth} 
\begin{center}
\caption{The sample of AGN host galaxies}
\label{tab:gal_specs}
\begin{tabular}{lllccc}
\hline
Galaxy	& $\alpha_{\mathrm{2000}}$& $\delta_{\mathrm{2000}}$& 	$z$	 &	Instrument   &	$\mathrm{T_{exp}}$ \\
	& (\textit{h m s}) & (\textit{d m s}) & &	& $(\mathrm{s})$ \\
\hline
Mrk 279  & 13 53 3.4 & +69 18 30  &  0.030451  & HexPak &     10800  \\
NGC 3227 & 10 23 30.6 & +19 51 54   & 0.003859	 & HexPak &   7200     \\
NGC 3516 &  11 06 47.5 & +72 34 07  & 0.008836	 & HexPak &   7200     \\
NGC 4051 & 12 03 9.6 & +44 31 53 & 0.002336	 & HexPak &   7200     \\
NGC 4151 & 12 10 32.6 & +39 24 21 & 0.003319	 & HexPak &   7200     \\
NGC 4253 & 12 18 26.5 & +29 48 46  & 0.012929	 & HexPak &  10800     \\
NGC 4593 & 12 39 39.4 & -05 20 39  & 0.009      & HexPak &   7200     \\
NGC 5548 & 14 17 59.5 &  +25 08 12 & 0.017175   & HexPak &  14400     \\
NGC 6814 & 19 42 40.6 & -10 19 25 & 0.005214  & WiFeS     &   27000  \\
Mrk 79   & 07 42 32.8 & +49 48 35 & 0.022189  &  NIFS     &  3300    \\
\hline

\end{tabular}
\end{center}
\textbf{Notes.} Column 1: galaxy name, Column 2: right ascension, Column 3: declination, Column 4: redshift as quoted on NED, Column 5: instrument with which data for this program were taken, Column 6: total exposure time.
\end{minipage}
\end{table*}
\subsubsection{Data Reduction} \label{subsubsec:hexdatred}
Spectra were reduced using standard procedures for multi-fiber spectrographs. Spectra were bias-subtracted, overscan-corrected and dark corrected using standard IRAF\footnote{IRAF is distributed by the National Optical Astronomy Observatory, which is operated by the Association of Universities for Research in Astronomy (AURA) under a cooperative agreement with the National Science Foundation.} tasks. Flat-fielding, wavelength calibration, fiber-to-fiber throughput corrections, flux calibration and sky subtraction were all done using the IRAF task DOHYDRA in the HYDRA package. Since windshake was a concern during observing it was necessary to carefully align exposures, which was done using maps of localized AGN emission. As part of this process, wherever possible we compared data taken when it was calm with data taken during periods of higher wind for the same galaxy, in order to gauge the effects. We found the effects to be minimal, and we do not expect it to significantly impact our results. Once aligned, individual exposures for each galaxy were median combined and cosmic ray rejection was carried out during this process. 

In all cases the final reduced spectra for fibers covering the central regions of each galaxy had sufficient signal-to-noise per pixel ($\mathrm{S/N}\gtrsim30$) for a reliable assessment of the stellar kinematics in the bulge. In most cases this was also true for fibers extending out into the disk, with the notable exception of Mrk 279, for which we were not able to collect enough signal to detect stellar absorption features beyond the central region of the galaxy. 

\subsubsection{Analysis} \label{subsubsec:hexanal}
Stellar kinematics were determined using the penalized pixel-fitting method (pPXF) of \cite{Cappellari04}. This method takes a set of stellar template spectra and convolves them with a line-of-sight velocity distribution (LOSVD) to find a best fitting model of the galaxy spectrum. With sufficiently high quality data, pPXF can be used to reliably determine up to the first six Gauss-Hermite coefficients of the LOSVD, though for the purposes of this work we are interested in only the first two (V and $\sigma$). 

The pPXF code will accept a variety of stellar template spectra, which can be fit individually or simultaneously to better represent the mix of stellar spectral types in the galaxy. It will also account for differences in the instrumental dispersion between template and target spectra, so libraries of velocity templates obtained with other instruments may be used in place of, or supplementary to, velocity templates observed as part of the science program. For this analysis we have used a variety of template spectra from the publicly available Elodie Stellar Library version 3.1 \citep{Prugniel07}. The Elodie library consists of high resolution spectra from a subsample of 1388 stars from the Elodie Archive, covering the wavelength range $4000-6800\mathrm{\AA}$. A variety of spectral types are available and, after some testing to identify which spectral types from the archive best fit our data, we chose a set of six giant stars ranging from G8-M0 to use as our template sample. 

In each case a multi-step process was used to fit the LOSVD; pPXF was initially run using all six template stars. The stars that contributed to the LOSVD in $\gtrsim50\%$ of the fibers were identified, and pPXF was run again with that subset to get our final V and $\sigma$ values. Finally, pPXF was run with each contributing template star individually, so that the spread in determined values could be used in error estimation. In all cases the region over which the spectra were fitted was restricted to a rest-frame wavelength range $\sim5130-5360\mathrm{\AA}$, containing the Mg\textit{b} and Fe ($5270,5335\mathrm{\AA}$) absorption lines. This effectively removed parts of the spectrum that may be contaminated by strong AGN emission lines. While continuum emission is certainly present in the fitted wavelength region, it does not overwhelm the host galaxy starlight in most spaxels, nor does it significantly affect the shape and width of the stellar absorption features, so deblending of the AGN light from the host galaxy spectra (as done by e.g.\ \citealt{Husemann16} when determining $\sigma_{\star}$ for a galaxy hosting a quasi-stellar object) was not found to be necessary. Those spaxels in which the AGN emission does dominate were not included in our final analysis (see Section \ref{sec:results}).

The CaII triplet is generally considered to set the benchmark for $\sigma_{\star}$ measurements, so there is some potential for bias in our measurements when compared to others in the literature. Comparison of $\sigma_{\star}$ determinations from fitting of CaII and Mg \textit{b} spectral regions by \cite{Barth02} showed that, in general, the Mg \textit{b} fitting is more sensitive to template mismatch, though this effect is less significant for late-type host galaxies with lower $\sigma_{\star}$ (as is predominantly the case in our sample). Our use of a range of templates to fit the LOSVD and determine the uncertainty, rather than relying on a single template stellar spectrum, should help mitigate this effect and account for it in the quoted error. Indeed a more recent study by \cite{Woo15}, who used pPXF to compare $\sigma_{\star}$ determinations from the two spectral regions for a sample of narrow-line Seyfert 1 galaxies, found that the measurements were completely consistent. Consequently, while we cannot rule out the possibility of some bias due to this effect, we do not expect it to be significant. 
\subsection{NGC 6814} \label{subsec:other_6814}

\subsubsection{Observations} \label{subsub:6814obs}
Similar to our sample of galaxies observed with HexPak, NGC 6814 is a nearby, low-inclination, late-type Seyfert galaxy with significant substructure including a bar. Details of the galaxy are shown in Table \ref{tab:gal_specs}. 
NGC 6814 was observed with the Wide-Field Spectrograph (WiFeS) in July 2012 under observing program 2120061. WiFeS is an optical, dual-beam, integral-field spectrograph mounted on the Australian National University (ANU) 2.3m telescope at the Siding Spring Observatory \citep{Dopita07,Dopita10}. It has a large contiguous $25\arcsec \times 38 \arcsec$ field of view, and is seeing limited with $1\arcsec \times 0\farcs5$ spaxels. The dual-beam construction of WiFeS allows it to operate across the complete optical wavelength range ($3300-9000\mathrm{\AA}$) with a single exposure in low-resolution mode ($\mathrm{R}=3000$), and two exposures in high-resolution mode ($\mathrm{R}=7000$). 

Observations targeting the Calcium triplet ($\lambda 8498, 8542, 8662 \mathrm{\AA}$) were made with the high-resolution I-band (I7000) grating. We also observed in the B-band (B3000 grating), however since those data did not yield sufficient S/N to detect stellar absorption features we do not discuss them further. Observations were made during bright time with a standard sky-object-object-sky pattern, interspersed with calibration arc-lamp, bias and flat frames, which were necessary to account for expected variations in the instrument throughout the night. This yielded 15 on-source exposures of 1800s each, for a total integration time of 27ks. Spectrophotometric standards were observed throughout each night to facilitate telluric correction and flux calibration.

\subsubsection{Data Reduction} \label{subsub:6814datred}
Data reduction was performed using the open-source, Python-based, PyWiFeS data reduction pipeline \citep{Childress14}. PyWiFeS makes use of standard Python libraries to perform rapid data-reduction and produce high-quality, fully processed spectra. 
 
Following standard reduction procedures for IFU spectroscopy, spectra are first bias-subtracted (using the bias frames taken nearest in time to the science frame), after which cosmic ray rejection is done for each slitlet. Spectra are flat-fielded, corrected for the illumination function and for atmospheric differential refraction, and then spatially calibrated. Flux calibration and telluric corrections are performed, and finally the spectra are wavelength calibrated and reformatted into datacubes. After the reduction pipeline is complete, datacubes are aligned and median combined with the IRAF task imcombine.   
\subsubsection{Analysis} \label{subsub:6814anal}
As for the HexPak data, stellar kinematics were determined using pPXF. In order to have sufficiently high S/N per pixel we restricted our analysis to spaxels within the central $15\arcsec \times 15\arcsec$, and rebinned along the x-axis to get square $1\arcsec$ spaxels. 

Template spectra were taken from the Near-IR Ca II Triplet Library \citep{Cenarro01}, which consists of high quality spectra of 706 stars of diverse spectral types. We initially used a sample of nine spectra consisting of seven G, K, and M giants as well as two main sequence stars (G0V and B8V). We found that the LOSVD of the galaxy was best modeled by a combination of a K3III star and a M6III star. 

\subsection{Mrk 79} \label{subsec:m79}
\subsubsection{Data and Reduction} \label{subsub:m79datred}
Data taken with Gemini-North's Near-Infrared Integral Field Spectrometer (NIFS) are available for Mrk 79 from the Gemini Observatory Archive\footnote{https://archive.gemini.edu/} \citep{NIFS}. NIFS provides for very high spatial resolution ($0\farcs1$ when used with their adaptive-optics system ALTAIR) and good spectral resolution over a small $3\arcsec \times 3 \arcsec$ field of view. The field of view of NIFS is significantly smaller than that of HexPak and WiFeS, so these data probe a smaller region of the galaxy than is probed in the other targets in this study. However the field of view does extend out to $\sim r_{e}/2$, which is large enough to give us insight into the bulge stellar dynamics and allow for a good determination of $\sigma_{\star}$.  

The archived data were originally obtained and published by \cite{Riffel13}. Details of the observations, as well as results of analysis of the gas dynamics, can be found in that paper so we provide only a brief overview of the pertinent details here. Observations were obtained using the ALTAIR adaptive-optics system in September 2010 under programme GN-2010B-Q-42. The $\mathrm{K}_{\l}$-band observations covered the spectral region $2.10 - 2.53\mu \mathrm{m}$ centered at $2.3\mu \mathrm{m}$, with a spectral resolution of $\mathrm{FWHM}\approx3.5 \mathrm{\AA}$ and a total integration time of 3300s.

Data reduction was based on sample reduction scripts provided by Gemini\footnote{Sample scripts and basics reduction methods are available at http://www.gemini.edu/sciops/instruments/nifs/data-format-and-reduction}, and followed the prescription laid out by Riffel et al., using tasks within the NIFS IRAF package as well as generic IRAF tasks. Images were trimmed, flat-fielded, sky subtracted, wavelength calibrated and spatially rectified. Telluric corrections were made, and a basic flux-calibration was accomplished by interpolating a black body function with the telluric standard star spectrum before the telluric correction was made. The individual datacubes were aligned and combined, during which process cosmic ray rejection was performed. The final datacube was trimmed to a $1\farcs6 \times 1\farcs6$ field of view, centered on the AGN.

\subsubsection{Analysis} \label{subsub:m79anal}
Since these data were originally obtained for analysis of the gas dynamics, there is not sufficient S/N in each spaxel to reliably identify stellar absorption features. Consequently we bin the spaxels to obtain spectra with S/N $\approx 150$, which we found to be necessary to be able to fit the LOSVD using pPXF. The Voronoi binning method of \cite{Cappellari03}, which allows for variable bin sizes, is used to bin the spectra to a constant S/N across the field of view. This method results in eight bins where stellar absorption features can be reliably identified and fit. Four bins in the center of the field are discarded because the AGN flux overwhelms all stellar features in the spectra.  

Velocity templates are taken from the library of stellar spectral templates made available by Gemini \citep{Winge09}. Since observations of these stars were made with NIFS, this library is ideal for use with these data. As with the analysis described previously, we start with a selection of G, K, and M giant stars and run pPXF to model the LOSVD for the binned galaxy data. We run pPXF a second time including only the template stars that are found to contribute significantly from the first run, and finally pPXF is run with each template star individually for error analysis. We found that the LOSVD was best fit by a combination of a K4III and M2III stars, with the K4III star dominating in most cases.  

\section{Results and Discussion} \label{sec:results}
\subsection{Bulge Effective Radii}

Accurate determinations of $r_{e}$ are available for all of the galaxies in our sample from \cite{Bentz09,Bentz13,Bentz16}, and are shown in Table \ref{tab:kin}. These $r_{e}$ are determined from detailed GALFIT \citep{Peng02,Peng10} models of surface brightness decompositions of Hubble Space Telescope (HST) images. Two sets of models are used in these analyses, both of which accurately isolate the AGN contribution from the galaxy surface brightness features. We make use of the simple models rather than the optimal models (see \citealt{Bentz13} for details of the differences) as they fit fewer components, accounting only for the PSF and the physical structures present in the galaxies (such as bulge, bars, and disks). In some cases the models fit two components to the bulge, in which case we take an average of the $r_{e}$ for each component, to obtain the value quoted in Table \ref{tab:kin}. Full details of the method are given in the relevant references.
\subsection{Stellar Kinematics} \label{subsec:maps}

\begin{figure*}
\begin{minipage}{\textwidth}
\begin{center}
\includegraphics[scale=1]{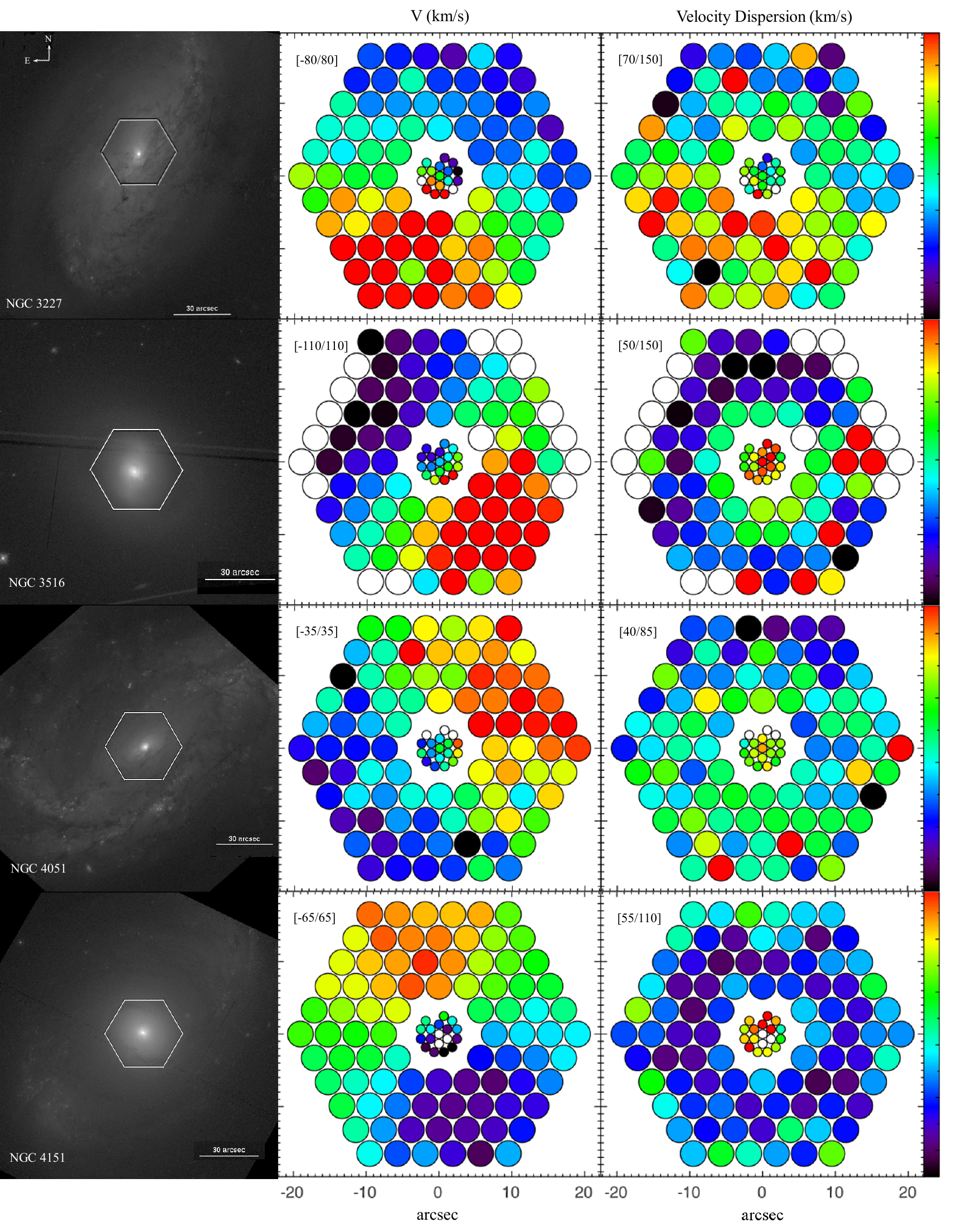}
\caption{From left to right: panel 1 shows an HST image of the galaxy with a scale bar in the lower right, the HexPak field of view indicated, and the galaxy name in the lower left, panels 2 and 3 show maps of the velocity and stellar velocity dispersion ($\sigma_{\star}$) respectively. Blank circles indicate fibers with insufficient S/N to resolve stellar absorption features. The range of plotted values is shown in the top left corner of each map.}
\end{center}
\end{minipage}
\label{fig:hexmaps_1}
\end{figure*}

\begin{figure*}
\begin{minipage}{\textwidth}
\begin{center}
\includegraphics[scale=1]{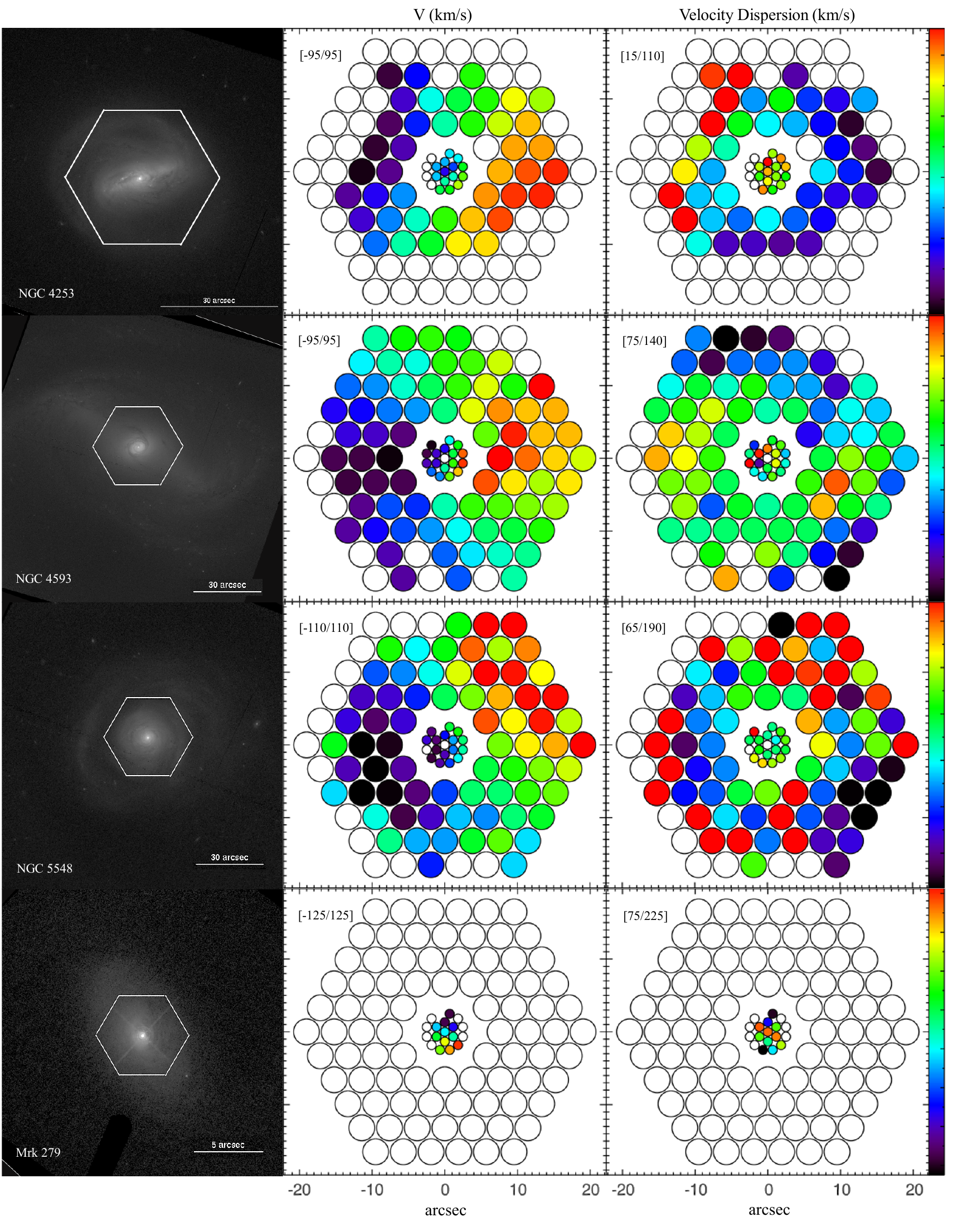}
\end{center}
\caption{The same as Figure \ref{fig:hexmaps_1}.}
\end{minipage}
\label{fig:hexmaps_2}
\end{figure*}

Kinematic maps of the first and second moments of the LOSVD, V and $\sigma_{\star}$, are shown for the HexPak sample in Figures \ref{fig:hexmaps_1} and \ref{fig:hexmaps_2}, along with HST images of each galaxy. As can be seen in the maps there is a gap between the central bundle and the outer fibers that means we do miss some information in relevant parts of each galaxy. However for six of the eight galaxies in the sample $\mathrm{r}_{e}<3\arcsec$, and is therefore completely covered by the central fiber bundle. For NGC 4593 and NGC 5548, $\mathrm{r}_{e}$ extends out to include the first two rings of outer fibers, so while the gap does impact our spatial resolution in these cases, the region within $\mathrm{r}_{e}$ is still well sampled. For Mrk 279 we only have sufficiently high S/N data to map the central region of the galaxy. However since this region extends beyond $r_{e}$ for the bulge of this galaxy (see below, and Table \ref{tab:kin}), it allows us to determine $\sigma_{\star}$ in the same manner as for the rest of the sample.

\begin{figure*}\vspace{-20ex}
\begin{minipage}{\textwidth}
\begin{center}
\includegraphics[scale=0.8]{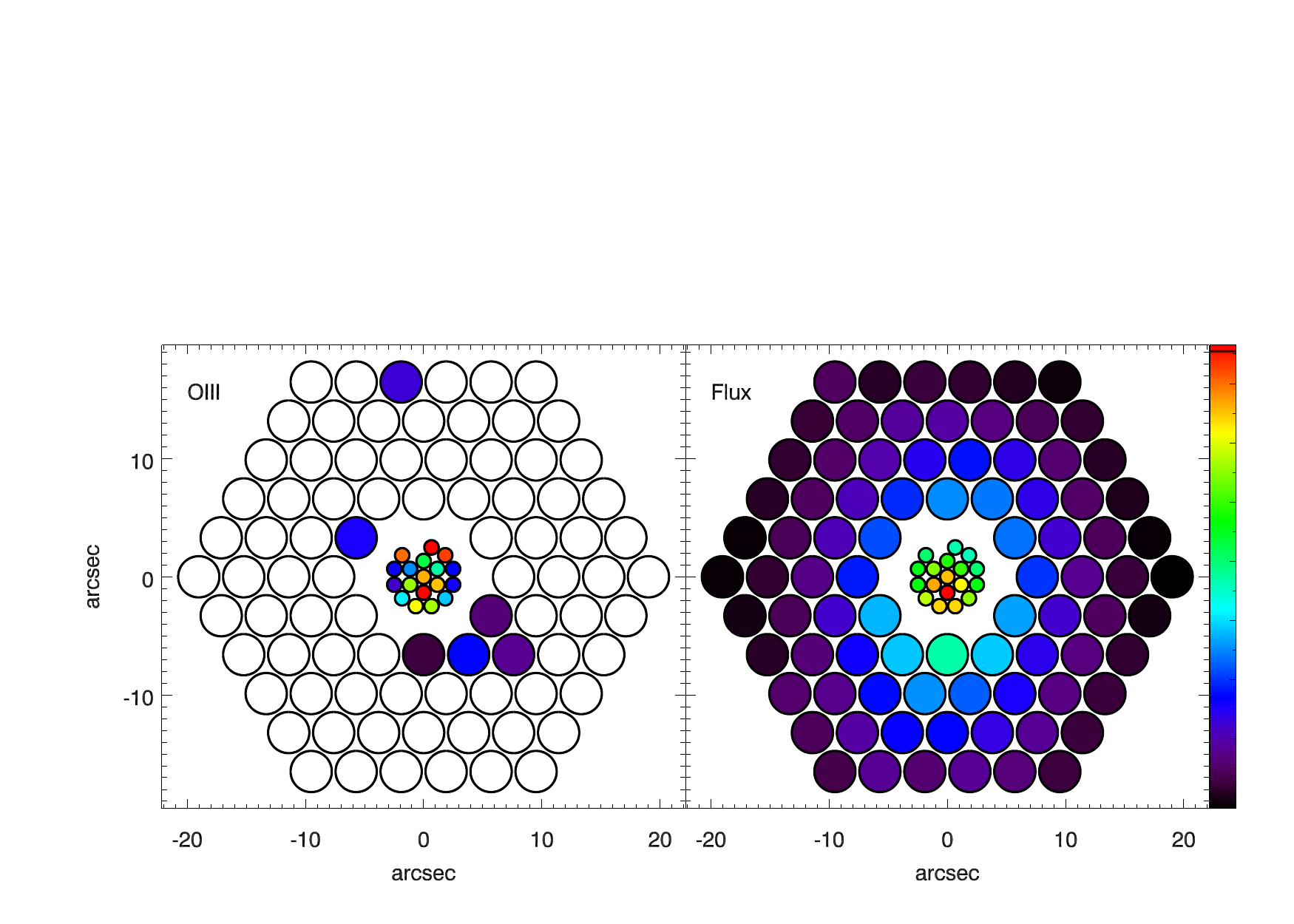}\vspace{-1ex}
\end{center}
\caption{\textbf{right panel:} Broadband flux reconstruction for NGC 3516, plotted on a logarithmic scale. The flux units are arbitrary and the scaling is set so that the darkest fibers are the lowest flux. As can be seen, the observations were made and aligned in such a way that the AGN is offset south from the center of the field by $1\arcsec$. \textbf{left panel:} Map of the OIII ($\lambda5007\mathrm{\AA}$) emission for NGC 3516. This map is consistent with the imaging of \cite{Schmitt03}, which showed an S-shaped emission region, extended in the north-south direction. Blank circles indicate fibers with no detectable emission.}
\end{minipage}
\label{fig:flux_3516}
\end{figure*}

As an example of the data quality Figure \ref{fig:flux_3516} shows image reconstructions for NGC 3516 from the broadband flux (right panel), and the narrow OIII ($\lambda5007\mathrm{\AA}$) emission. OIII imaging of NGC 3516 by \cite{Schmitt03} showed an extended S-shaped emission region oriented primarily north-south (see the top right panel of their Figure 10), which is consistent with what we see here.

\begin{figure*}
\begin{minipage}{\textwidth}
\begin{center}
\includegraphics[scale=1]{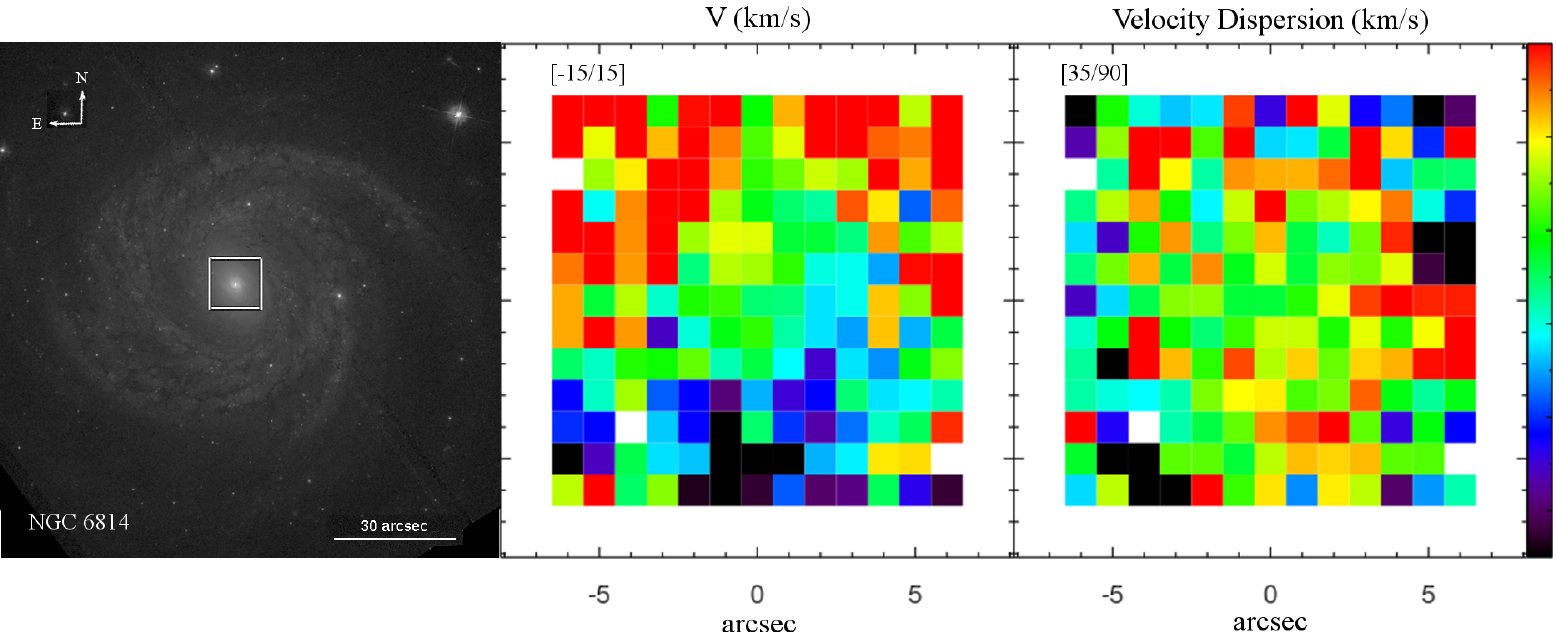}
\end{center}
\caption{The same as Figure \ref{fig:hexmaps_1} for NGC 6814, with the WiFeS field of view shown on the image. Blank pixels indicate those for which it was not possible to reliably identify stellar absorption features.}
\end{minipage}
\label{fig:6814_maps}
\end{figure*}

\begin{figure*}
\begin{minipage}{\textwidth}
\begin{center}
\includegraphics[scale=1]{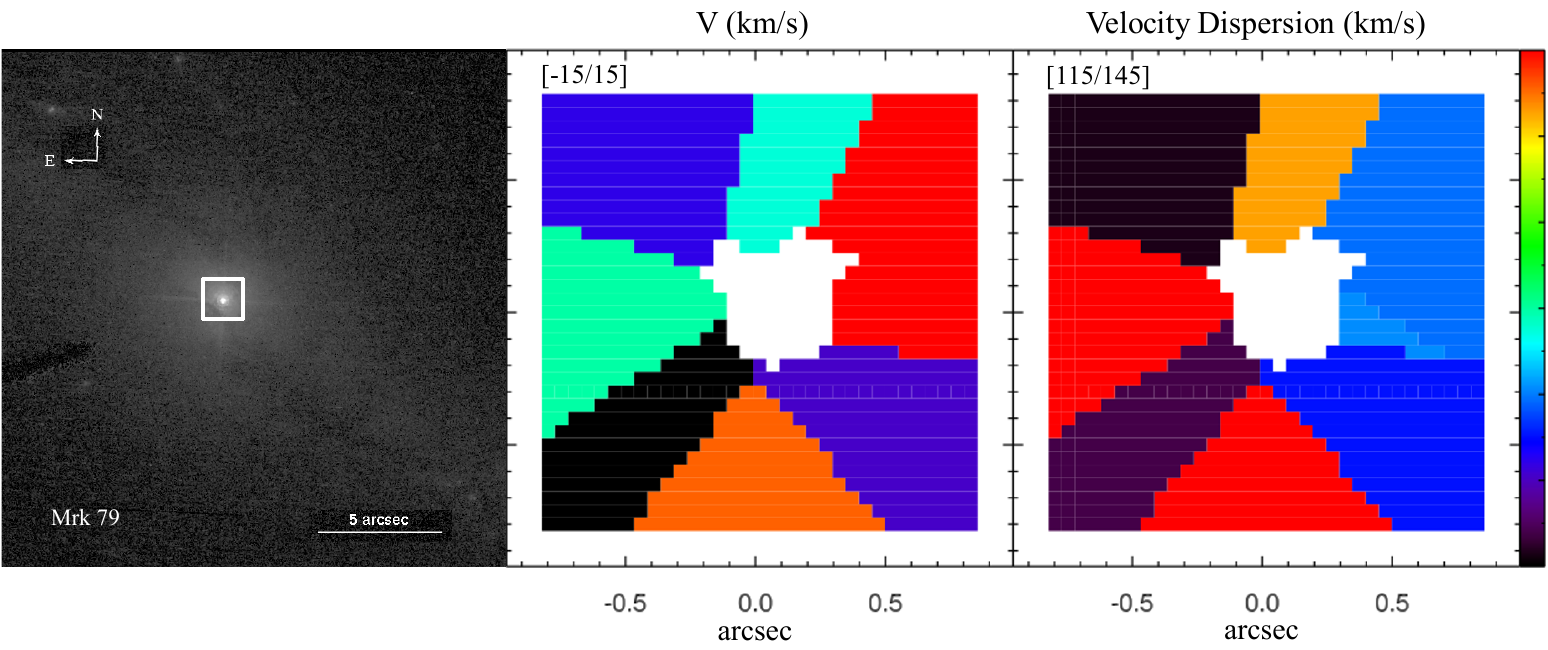}
\end{center}
\caption{The same as Figure \ref{fig:hexmaps_1} for Mrk 79, with the NIFS field of view shown on the image. Values for the central bins (approximately the central $0\farcs5$) are not plotted since stellar absorption features could not be reliably identified.}
\end{minipage}
\label{fig:mrk79_maps}
\end{figure*}

Figures \ref{fig:6814_maps} and \ref{fig:mrk79_maps} show the kinematic maps and images for NGC 6814 and Mrk 79. For NGC 6814, as for the HexPak sample, $\mathrm{r}_{e}$ is well covered by the WiFeS field of view. For Mrk 79 the central $\sim 0\farcs5$ are too heavily contaminated by AGN emission to be able to reliably identify stellar absorption features, so those bins have been removed. The NIFS field of view is not sufficient to cover $r_{e}$ for Mrk 79, however with a radius of $0\farcs85$ it does extend nearly to $r_{e}/2$. 

The full set of $\sigma_{\star}$ maps shows clearly the value of spatial resolution. Variation is apparent throughout the bulge of each galaxy, likely indicative of contamination from the disk and from the significant substructure that is visible in the HST images. 

Table \ref{tab:kin} gives $\mathrm{r}_{e}$ and average $\sigma_{\star}$ within $\mathrm{r}_{e}$ for each galaxy (columns 2 and 4). In every case $\sigma_{\star}$ was determined by taking a weighted average of $\sigma_{\star}$ values for each spaxel (or bin, in the case of Mrk 79) within $\mathrm{r}_{e}$. In cases where spaxels only partially cover the region within $r_{e}$ (i.e. including the spaxel gives an aperture larger than $r_{e}$, but excluding it gives an aperture that is smaller) then the spaxel is additionally weighted according to the fraction that should be included.

For Mrk 79 a standard aperture correction has been applied following \citealt{vandenbosch16}:
\begin{equation}\label{eq:apcor}
\dfrac{\sigma_{\star_{e}}}{\sigma_{\star}}=\left(\dfrac{r_{NIFS}}{r_{e}}\right)^{0.08}
\end{equation}
to give $\sigma_{\star}$ within $\mathrm{r}_{e}$. The table entry for Mrk 79 gives the corrected value of $\sigma_{\star}$ with the uncorrected value in parentheses. 
Estimated errors are the sum in quadrature of the fitting error from pPXF, and the standard deviation in $\sigma_{\star}$ among the values determined when pPXF was run with each contributing template individually (see section \ref{subsubsec:hexanal}). 

\begin{table*}
\begin{minipage}{\textwidth}
\begin{center}
\caption{Host galaxy $r_{e}$ and kinematics for our sample}
\label{tab:kin}
\begin{tabular}{ccccccc}
\hline
\hline
Galaxy  & $\mathrm{r}_{e}\:(\arcsec)$& Refs.  & $\sigma_{\star}\:\mathrm{(km\:s^{-1})}$  & Standard Deviation ($\mathrm{km\:s^{-1}}$) & $\sigma_{\star_{lit}}\:\mathrm{(km\:s^{-1})}$& Refs.  \\
\hline
Mrk 279     & $1.6$ & 1 & $153 \pm 7 $ & 26  & $197 \pm 12 $& 4 \\
NGC 3227    & $2.7$ & 3 & $114 \pm 3 $ & 13 & $92 \pm 6 $& 6\\
NGC 3516    & $2.1$ & 3 & $139 \pm 4 $ & 12  & $181 \pm 5 $& 4\\
NGC 4051    & $1.0$ & 3 & $74 \pm 2 $ & 4    & $89 \pm 3 $& 4\\
NGC 4151    & $2.1$ & 3 & $105 \pm 5 $ & 15   & $97 \pm 3 $& 4\\
NGC 4253    & $1.4$ & 2 & $84 \pm 4 $ & 9    & $93 \pm 32 $& 5\\
NGC 4593    & $11.5$ & 3 & $113 \pm 3 $ & 14  & $135 \pm 6 $& 4\\
NGC 5548    & $11.2$ & 3 & $131 \pm 3 $ & 34  & $195 \pm 13 $& 5\\       
NGC 6814 & $1.7$ & 2 & $ 71 \pm 3 $ &  5  & $ 95\pm 3 $ & 5 \\
Mrk 79   & $2.0$ & 1 & $ 120\:(129)\pm 9 $ & 21 & $130 \pm 12 $ & 4 \\ 
\hline
\end{tabular}
\end{center}
\textbf{References.} (1) \citealt{Bentz09}; (2) \citealt{Bentz13}; (3) \citealt{Bentz16}; (4) \citealt{Nelson04}; (5) \citealt{Woo10}; (6) \citealt{Woo13} \\
\textbf{Notes.} Column 1: galaxy name, Column 2: bulge $\mathrm{r}_{e}$, Column 3: reference for the value in column 2, Column 4: $\sigma_{\star}$ within $\mathrm{r}_{e}$ from our data, with associated $1\sigma$ uncertainty, Column 5: standard deviation for the set of $\sigma_{\star}$ values averaged to determine the value in column 4, Column 6: $\sigma_{\star}$ from the literature, Column 7: reference for the value in column 6.
\end{minipage}
\end{table*}

\subsection{Comparison with the Literature}
Table \ref{tab:kin} lists frequently used values of $\sigma_{\star}$ for each of the galaxies included in our sample, for comparison with our results. Some specific notes on how the literature values were determined for each galaxy are given below.

\subsubsection{Notes on Specific Galaxies}
Six of the ten quoted literature values are from the work of \cite{Nelson04}, for which the Ritchey-Chr{\`e}tien (RC) spectrograph on the Kitt Peak 4m Telescope was used with a $1\arcsec$ slit, targeting the CaII triplet. In all of these cases a single integrated spectrum was used, and the extraction apertures are given below.\\
\textit{Mrk 279:} An extraction aperture of $4\farcs7$ was used (compared with $r_{e}=1\farcs6$, used in this work).\\
\textit{NGC 3516:} An extraction aperture of $4\farcs3$ was used (compared with $r_{e}=2\farcs1$).\\
\textit{NGC 4051:} An extraction aperture of $5\farcs6$ was used (compared with $r_{e}=1\farcs0$).\\
\textit{NGC 4151:} An extraction aperture of $6\farcs5$ was used (compared with $r_{e}=2\farcs1$)\\
\textit{NGC 4593:} An extraction aperture of $6\farcs5$ was used (compared with $r_{e}=11\farcs5$).\\
\textit{Mrk 79:} An extraction aperture of $4\farcs1$ was used (compared with $r_{e}=2\farcs0$).\\
\textit{NGC 3227:} \cite{Woo13} used the near-IR spectrograph Triplespec at the Palomar Hale 5m telescope. They used a $1\arcsec$ slit and spatially resolved along the slit in order to correct for rotational broadening out to $r_{e}$, which they determined to be $3\farcs4$ (compared with $r_{e}=2\farcs7$). \\
\textit{NGC 4253:} \cite{Woo10} obtained H-band IFU spectra with OSIRIS on the Keck-II telescope. A single integrated spectrum was used to determine $\sigma_{\star}$, within an extraction aperture $1\farcs3 \times 3\farcs4$ (compared with $r_{e}=1\farcs4$).\\
\textit{NGC 5548:} \cite{Woo10} used the Double Spectrograph (DBSP) at the Palomar Hale 5m telescope with a $2\arcsec$ slit at $59^{\circ}$, targeting the CaII triplet. A single integrated spectrum was extracted, with a typical extraction radius of $2\arcsec-3\arcsec$ for the full sample of galaxies in their study (compared with $r_{e}=11\farcs2$). \\
\textit{NGC 6814:} As for NGC 5548, \cite{Woo10} used the DBSP with a $2\arcsec$ slit at $0^{\circ}$, and a $2\arcsec-3\arcsec$ extraction radius to obtain a single integrated spectrum (compared with $r_{e}=1\farcs7$). \\

\subsubsection{Discussion}
For NGC 4151, NGC 4253, and Mrk 79 our results are completely consistent with the literature values. However in the cases of other objects our values differ quite significantly from the literature and, excepting NGC 3227 and NGC 4151, are consistently lower. The most extreme difference is with NGC 5548, for which our calculated value is almost $60\:\mathrm{km/s}$ lower than that of \cite{Woo10}, with large discrepancies also evident for NGC 3516 and Mrk 279.   

There are several possible reasons for this\footnote{A useful discussion of these issues is presented by \cite{Kormendy13} in a note on Table 3 in Section 5}: (i) the fraction of $\mathrm{r}_{e}$ within which $\sigma_{\star}$ is reported varies in the literature, with $\mathrm{r}_{e}$ and $\mathrm{r}_{e}/8$ both commonly in use, as well as values in between. Since $\sigma_{\star}$ is not expected to be constant throughout the bulge, especially when kinematic substructure is present, this may have a significant impact. While it is possible to mitigate this somewhat by applying an aperture correction, such a correction makes certain assumptions about the galaxy which may not be valid. Perhaps more importantly, when investigating correlations between the central super-massive black hole and properties of the host galaxy, it is not known which is the best value to use in terms of being most physically meaningful or in order to achieve tighter correlations. IFU data can be used to study the impact of the chosen radius on the measured $\sigma_{\star}$, however such a study is beyond the scope of this work and will be part of a future paper.

(ii) There is not complete agreement on the mathematical definition of $\sigma_{\star}$ which is best to use. In this work we have chosen to use the second moment of the LOSVD, which is more commonly used in observational studies, but often a contribution from the rotation velocity is also included which will systematically broaden the observed value of $\sigma_{\star}$. A detailed comparison between the different definitions of $\sigma_{\star}$, similar to that done by \cite{Bennert15} with long-slit spectra, will be part of a future paper. 

(iii) The literature values for all but one of these galaxies have been determined from long-slit spectroscopy, rather than IFU spectroscopy. In comparing results from long-slit spectroscopy to IFU kinematics from the SAURON/ATLAS3D teams \citep{Emsellem07,Cappellari13}, \cite{Kormendy13} find that $\sigma_{\star}$ tends to be smaller when determined from the IFU data (see their Figure 11 and their note for Table 3), which is consistent with our results. In investigating stellar velocity dispersion estimates from long-slit spectroscopy, both \cite{Kang13} and \cite{Woo13} compared results from single extraction apertures of various sizes with results from small extraction windows along the slit (i.e. spatially resolving along the slit, see also \citealt{Harris12}). They found that, for disk galaxies, rotational broadening tended to cause $\sigma_{\star}$ to be over-estimated by as much as $20\%$ for spectra extracted from a single aperture, compared with spatially resolved spectra, and that this effect depends strongly on the size of the extraction aperture (though it was detectable in all apertures tested), the inclination of the disk, and the maximum rotational velocities. The spatial resolution of our IFU spectra means that our $\sigma_{\star}$ determinations should be less susceptible to the effects of disk contamination identified in these studies. However it is worth pointing out that NGC 3227 is one of the galaxies that \cite{Woo13} used for their study, and the quoted literature $\sigma_{\star}$ value in Table \ref{tab:kin} for NGC 3227 is their corrected value.
 
It is clear from the maps of $\sigma_{\star}$ that long-slit spectra may result in significantly different fits to the LOSVD, depending on the position angle of the slit and the number of different slit positions included in the analysis. The presence of significant kinematic substructure, which may be visible within the aperture and impossible to distinguish from bulge light, has been shown to be particularly problematic when trying to determine a bulge $\sigma_{\star}$ (see discussion in Section \ref{sec:intro}). 

\begin{figure*}
\begin{minipage}{\textwidth}
\begin{center}
\vspace{2ex}
\includegraphics[scale=0.6]{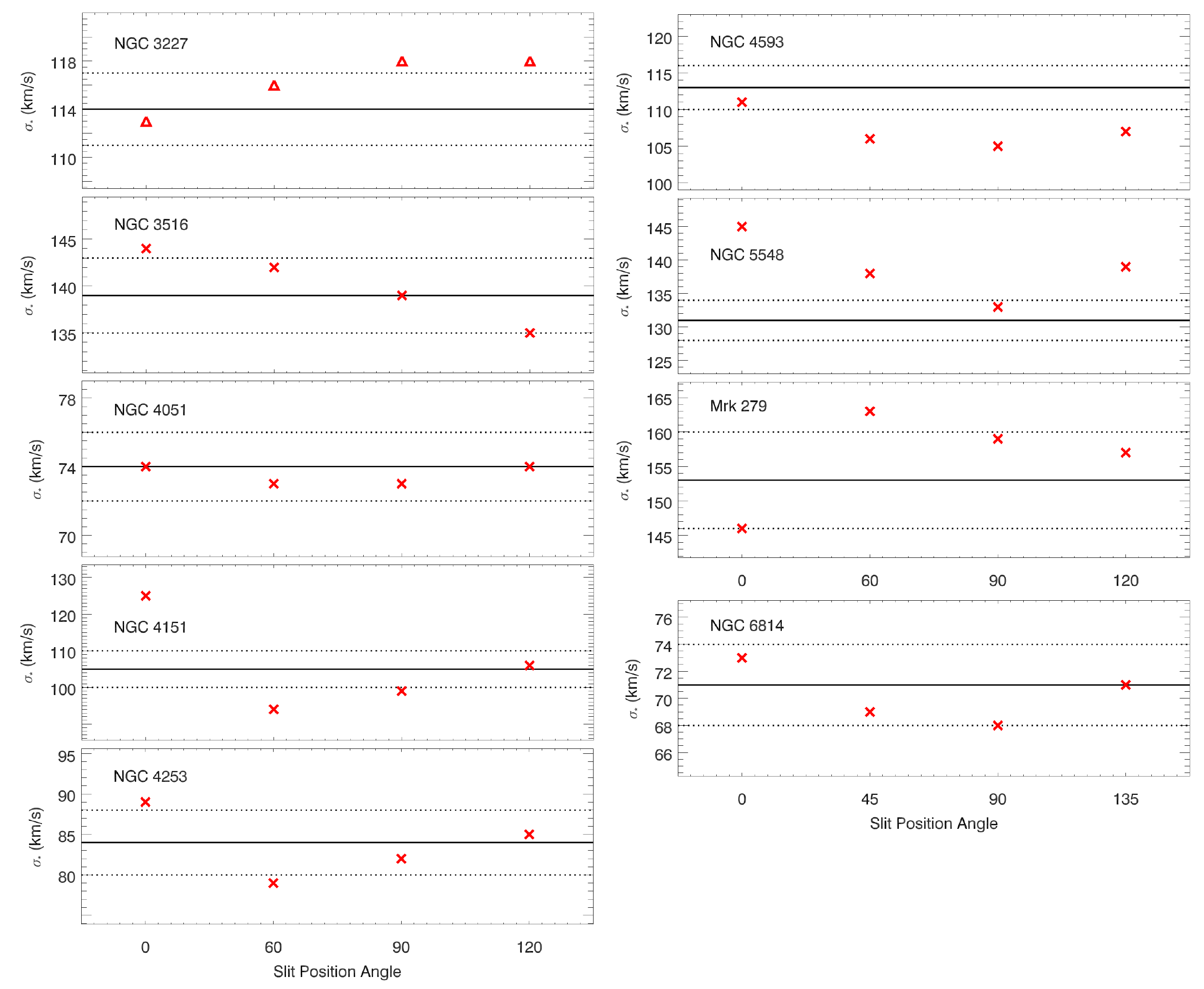}
\caption{Plots of $\sigma_{\star}$ as a function of position angle for simulated $1\arcsec$ slits at four different positions; $0^{\circ}$, $60^{\circ}$, $90^{\circ}$, $120^{\circ}$ for galaxes observed with HexPak, and $0^{\circ}$, $45^{\circ}$, $90^{\circ}$, $135^{\circ}$, for NGC 6814 (the difference is due to the geometry of the IFU). The slit extends to $r_{e}$ for each galaxy. The $\sigma_{\star}$ values from Table \ref{tab:kin} for each galaxy are indicated by a solid line on each plot, with dotted lines showing the $1\sigma$ uncertainties. Galaxy names are given in the top left corner of each plot.}
\label{fig:PAplots1}
\end{center}
\end{minipage}
\end{figure*}

In order to better illustrate this last problem, Figure \ref{fig:PAplots1} shows plots of $\sigma_{\star}$ within $r_{e}$, for simulated $1\arcsec$ slits at four different position angles, for all of the galaxies in the sample except Mrk 79 (for which a meaningful comparison cannot be made, given the spatial resolution and field of view of the data). It is clear from these plots that $\sigma_{\star}$ can vary significantly with PA, and in many cases this variation is larger than the quoted error in our determined $\sigma_{\star}$ values (our quoted $\sigma_{\star}$ and error are indicated by the solid and dotted lines, respectively, in each plot). The most significant variation occurs for NGC 4151, with a total difference of $31\:\mathrm{km\:s^{-1}}$. In contrast, all slit measurements for NGC 4051 are consistent with each other and with our quoted measurement.

\begin{figure}
\begin{center}
\vspace{2ex}
\includegraphics[scale=0.5]{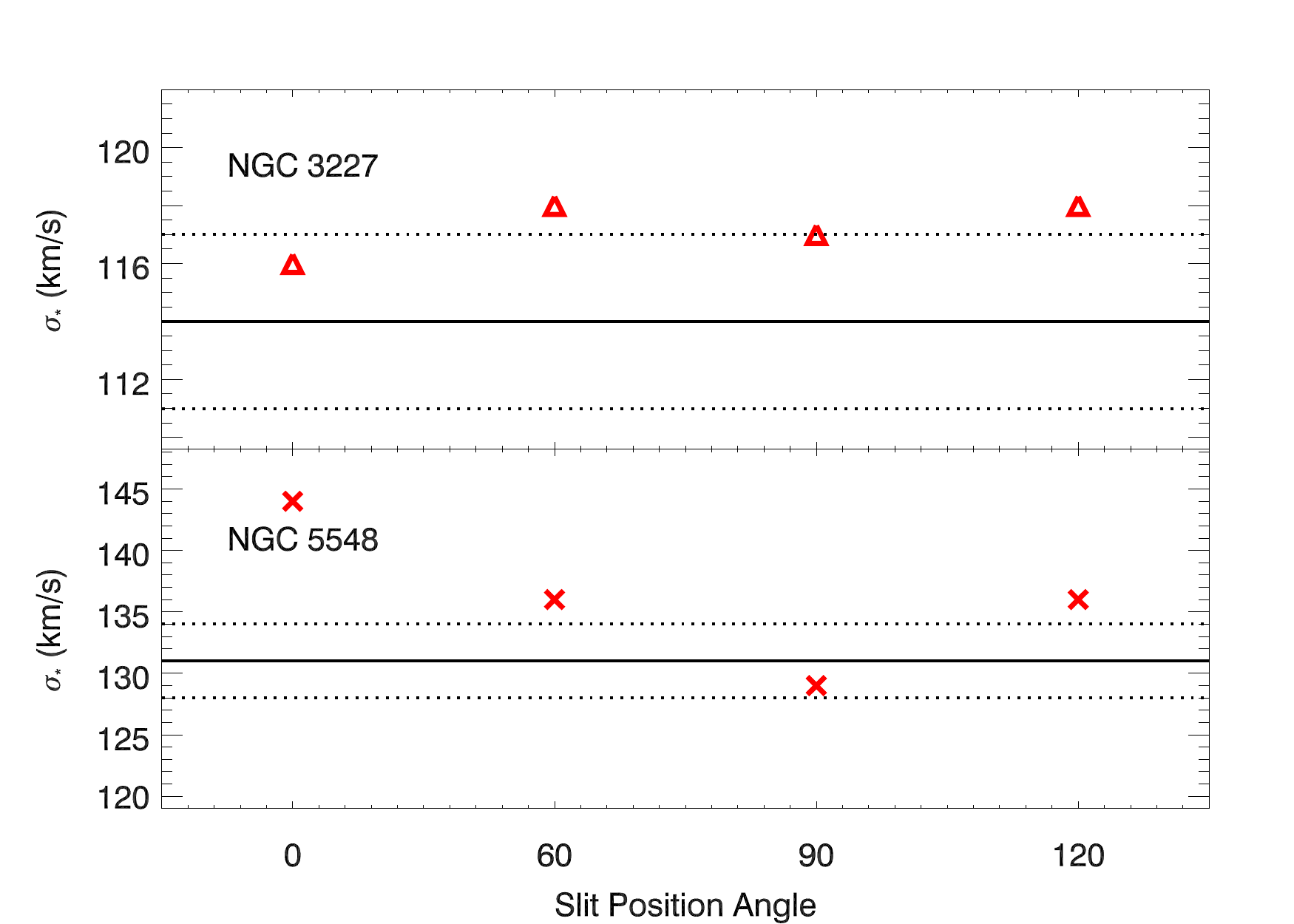}
\caption{Plots of $\sigma_{\star}$ as a function of position angle for NGC 3227 and NGC 5548, similar to Figure \ref{fig:PAplots1} but where the slit extends to the edge of the field of view ($r\approx18\arcsec$).}
\label{fig:PAplots2}
\end{center}
\end{figure}

Since many studies use extraction apertures larger than $r_{e}$, Figure \ref{fig:PAplots2} provides a comparison with Figure \ref{fig:PAplots1} for two galaxies, NGC 3227 and NGC 5548, with the slits extending to $r\approx18\arcsec$. These plots are preliminary and will be provided for the full sample, along with a detailed analysis, in a future paper. However, for now it is interesting to note that the values for NGC 3227 in Figure \ref{fig:PAplots2} are all slightly higher than those in Figure \ref{fig:PAplots1}, while the opposite is true for NGC 5548. 


In addition to the statistical error, column 5 of Table \ref{tab:kin} gives the standard deviation among the spaxel values that have been averaged together to produce the quoted $\sigma_{\star}$ for each galaxy. We include this value in an effort to quantify somewhat the variation in $\sigma_{\star}$ across $\mathrm{r}_{e}$ for each galaxy. While this does not reflect a measurement error or indicate how well the LOSVD is fitted for our spectra, it does give an indication of how variable $\sigma_{\star}$ is, and therefore it may also be a valuable indicator of how well constrained $\sigma_{\star}$ is for use in fitting the $M_{BH}-\sigma_{\star}$ relation. 

We present an updated calibration of the $M_{BH}-\sigma_{\star}$ relation for AGN, based in part on these results, in \cite{Batiste16}. This re-calibration is based exclusively on a sample of galaxies with secure $\sigma_{\star}$ determinations from IFU observations, and well constrained VPs from reverberation-mapping. The sample of galaxies included spans 3 orders of magnitude in $M_{BH}$, so the $M_{BH}-\sigma_{\star}$ relation is well sampled. The new fit allows for a recalculation of the scale factor $f$ that calibrates black hole masses from reverberation mapping. 

\section{Summary}

We have presented the results of two observing programs with which we obtained integral-field spectroscopy for nine AGN host galaxies, all of which have secure black hole mass determinations from reverberation mapping. Along with data from the Gemini Observatory archive for Mrk 79, we have presented spatially resolved kinematics maps of the first and second moments of the LOSVD for ten active galaxies. In all but one case these maps probe well beyond $r_{e}$, and in the case of Mrk 79 the field of view extends to $\sim r_{e}/2$, so the stellar dynamics of the bulge are well sampled in all cases. The maps (Figures \ref{fig:hexmaps_1} - \ref{fig:mrk79_maps}) show how $\sigma_{\star}$ varies within the field of view, possibly indicating the effects of dynamically distinct substructure, and highlighting the need for spatial resolution when trying to constrain bulge dynamics. To illustrate the need for spatial resolution in the context of long-slit spectroscopy, we take two galaxies as an example and provide comparisons of $\sigma_{\star}$ determinations from four simulated slits at different position angles. These comparisons show clearly that any estimate of $\sigma_{\star}$ from long-slit spectroscopy will depend on the orientation of the slit on the galaxy. 

We present new $\sigma_{\star}$ determinations within $r_{e}$ for the ten galaxies in the sample, based on our IFU observations and the most recent and accurate estimates of $r_{e}$. In general our $\sigma_{\star}$ are lower than those previously found, although in three cases our results are consistent with previous estimates, and in the case of NGC 3227 our value is higher.

In this paper we have presented the first results of the \textbf{BRAVE} Program. A recalibration of the $M_{BH}-\sigma_{\star}$ relation for AGN based on these results is the subject of a separate paper. In addition, much more detailed investigation of bulge dynamics will be the subject of future papers, including separating dynamically distinct substructure and isolating the bulge, detailed study of the variation with position angle and extraction window for simulated slits and large fibers, and correction factors that may be applied when IFU data are not available. 

\section*{Acknowledgments}
MCB gratefully acknowledges support from the NSF through CAREER grant
AST-1253702 to Georgia State University. This work is based on observations at Kitt Peak National Observatory, National Optical Astronomy Observatory (NOAO Prop. ID:2015A-0199; PI:M. Batiste), which is operated by the Association of Universities for Research in Astronomy (AURA) under a cooperative agreement with the National Science Foundation, and on observations with the NASA/ESA Hubble Space Telescope. 
\bibliography{ref}

\end{document}